\documentclass[12pt]{article}

\usepackage{amssymb}
\usepackage{amsmath}
\usepackage{amscd}
\usepackage{latexsym}
\usepackage{graphicx}

\usepackage{caption}
\usepackage[symbol]{footmisc}

\usepackage{cite}

\newcommand{\be}{\begin{equation}}
\newcommand{\ee}{\end{equation}}
\newcommand{\Dlt}{\Delta}
\newcommand{\dlt}{\delta}

\newcommand{\br}{{\bf r}}
\newcommand{\ba}{{\bf a}}
\newcommand{\bk}{{\bf k}}

\newcommand{\vp}{\varphi}
\newcommand{\ep}{\varepsilon}
\newcommand{\al}{\alpha}
\newcommand{\ra}{\rightarrow}
\newcommand{\sgm}{\sigma}

\newcommand{\gm}{\gamma}
\newcommand{\om}{\omega}

\newcommand{\dgr}{\dagger}
\newcommand{\lbd}{\lambda}

\newcommand{\rgl}{\rangle}
\newcommand{\lgl}{\langle}
\newcommand{\cH}{{\cal H}}
\newcommand{\cL}{{\cal L}}

\newcommand{\cF}{{\cal F}}

\begin{document}

\begin{center}
{\Large{\bf Ordering in statistical systems on the way to the thermodynamic limit} \\ [5mm]

V.I. Yukalov$^{1,2,}$\footnote[1]{Corresponding author e-mail: yukalov@theor.jinr.ru} 
and  E.P. Yukalova$^3$} \\ [3mm]


{\it $^1$Bogolubov Laboratory of Theoretical Physics, \\
Joint Institute for Nuclear Research, Dubna 141980, Russia \\ [3mm]
                                     
$^2$Instituto de Fisica de S\~{a}o Carlos, Universidade de S\~{a}o Paulo, CP 369,  \\
 S\~{a}o Carlos 13560-970, S\~{a}o Paulo, Brazil}  \\ [3mm] 

{\it $^3$Laboratory of Information Technologies, \\
Joint Institute for Nuclear Research, Dubna 141980, Russia}
\end{center}

\vskip 3cm

\begin{abstract}
It is well known that the mathematically accurate description of ordering and related
symmetry breaking in statistical systems requires to consider the thermodynamic limit.
But the order does not appear from nowhere, and yet before the thermodynamic limit is 
reached, there should exist some kind of preordering that appears and grows in the 
process of increasing the system size. The quantitative description of growing order, 
under the growing system size, is developed by introducing the notion of {\it order indices}. 
The rigorous proof of the phase transition existence is a separate difficult problem that
is not the topic of the present paper. We illustrate the approach resorting to several 
models in the mean-field approximation, which makes it possible to demonstrate the notion 
of order indices for finite systems in a clear way. We show how the order grows on the 
way to the thermodynamic limit for Bose-Einstein condensation, arising superconductivity, 
magnetization, and crystallization phenomena.        
\end{abstract}

\vskip 1cm
{\parindent=0pt
{\bf Keywords}: Statistical systems, Thermodynamic limit, Order indices, Reduced density 
operators, Correlation operators  }

\newpage

\section{Introduction}

Different thermodynamic phases are usually characterized by order parameters related
to the specific symmetries of the phases. The order parameters, as is well known, 
become meaningful only in the thermodynamic limit, when the number of particles in 
the system and its volume tend to infinity. Only in this limit, it is possible to give
a mathematically correct description of spontaneous symmetry breaking and the related
phase transitions. For example, it is straightforward to show that, if a spin Hamiltonian 
possesses rotational or reversal symmetry, the average spin, hence magnetization, is 
identically zero for a finite system. Similarly, if a Hamiltonian enjoys global gauge
symmetry, Bose-Einstein condensation cannot appear in a finite system. In the same way,
when a Hamiltonian is translationally invariant, equilibrium crystalline state cannot arise 
in a finite system. Detailed discussion and mathematically accurate description of these 
facts can be found in the books 
\cite{Bogolubov_1,Bogolubov_2,Kardar_52,Bogolubov_3,Yukalov_4,Dupuis_63}. 

In order to describe phase transitions and spontaneous symmetry breaking in mathematically
accurate way, Bogolubov \cite{Bogolubov_1,Bogolubov_2,Bogolubov_3} developed the method 
of quasi-averages, where the Hamiltonian symmetry is broken by an additional term that 
is removed after the thermodynamic limit. It is important to stress that the 
{\it thermodynamic limit goes first}, and the symmetry-breaking term is removed only 
{\it after} that. In principle, it is possible to define the symmetry-breaking term so 
that it would go to zero almost simultaneously with, but anyway a bit slower than the 
thermodynamic limit \cite{Yukalov_6}. In any case, the thermodynamic limit is compulsory 
for the mathematically rigorous definition of order parameters.   

A principal question that has remained unclear is wether the ordering could happen before 
the thermodynamic limit, when the system is already sufficiently large, although not yet 
infinite, and how the order could grow under the increasing system size until the 
thermodynamic limit. It is the aim of the present paper to suggest a mathematical 
description for the process of order growth under the increasing system volume up to the 
thermodynamic limit. 

In order to avoid misunderstanding, let us specify what is the main picture we keep in mind 
and what are the secondary details that can be neglected and that we will not touch. We
consider a system of $N$ particles in volume $V$, hence density $\rho = N/V$, under fixed 
thermodynamic and external parameters. According to the ideology of the thermodynamic limiting
procedure, we then consider the system of a larger number of the same kind of particles in 
a larger volume, such that the density remains constant and all thermodynamic parameters
are the same, being fixed. In that way, we consider not a single system, with increasing 
volume, but a sequence of systems with increasing sizes keeping the same fixed density 
and thermodynamic parameters. This is exactly the procedure accepted in the thermodynamic 
limit.
  
We consider statistical systems with large numbers of particles $N \gg 1$. Because of this
inequality, although the number $N$ can be finite, but, due to its large value, boundary 
effects can be neglected, especially when $N \ra \infty$. Therefore we do not need to 
resort to the thermodynamics of small systems of the size of single molecules \cite{Hill_53}.
Since, we increase the system size explicitly, we do not need to involve finite-size scaling
\cite{Privman_54} that is used in numerical calculations, when one tries to extrapolate to 
an infinite system size the measurements accomplished for small systems. 

Of course, sometimes, when considering finite systems of intermediate size, one needs to 
take account of surface effects. For instance, the properties of finite clusters of magnetic 
materials can depend on surface properties, inducing, e.g., magnetic anisotropy 
\cite{Kaneyoshi_1991}. 

In certain situations, finite-size effects can play a role. For example, even versus odd
number of fermions can play a role for atomic nuclei and metallic clusters. Shape and 
orientation transitions can occur for small systems in external fields or under rotation
\cite{Birman_51}. Anisotropy in geometry is important for studying one- and two-dimensional
limiting cases \cite{Imry_1969}. However, we keep in mind the standard understanding of 
large systems as those for which boundary effects, although could yield some small corrections,
but are not important for the considered phenomena, so that to a good approximation can be 
neglected when $N \gg 1$ \cite{Lavis_2021}. We assume that there are no external fields 
imposing anisotropy throughout the whole system. 

We do not aim of studying the behavior of thermodynamic characteristics in the close 
vicinity of critical points. Therefore we do not consider the semi-phenomenological droplet 
approximation \cite{Fisher_55,Fisher_56,Harris_57} for describing the critical region. In 
droplet models one considers large systems as composed of many small droplets, with the 
energy and entropy of each of them described by the so-called leptodermous expansions in 
powers of the droplet volume, with phenomenological coefficients. In that way, a droplet 
approximation treats a large single system as being composed of numerous small droplets. 
This picture does not have much relation to our aim of considering a sequence of systems 
whose volume increases tending to the thermodynamic limit.     

Each physical system, under the fixed density and other thermodynamic parameters, is assumed 
to be equilibrium. In an infinite system, this can be an absolute equilibrium  while in finite 
large systems it is rather a metastable equilibrium. A mathematically rigorous characterization
of metastable states can be found in the review by Sewell \cite{Sewell_58}. Being interested
in equilibrium states, we do not touch nonequilibrium phenomena related to the nucleation and
temporal formation of a new thermodynamic phase or a new structure, starting from a state of 
metastability \cite{Abraham_59,Debenedetty_60,Pruppacher_61,Oxtoby_62}.     

Thus, we consider a sequence of equilibrium statistical systems differing from each other 
only by the size, that is by the number of particles and system volume, with fixed density 
and all thermodynamic and external parameters. Each system consists of a large number of 
particles, such that surface effects can be neglected. Our aim is to develop a quantitative
characterization of the evolving order under the increasing system size on the way to the
thermodynamic limit.     

The guiding idea for developing a mathematical approach that could describe the growing 
order under the increasing system size can be qualitatively characterized as follows. 
Order, emerging in a system, depends on the relation between an effective correlation 
length and the system size. The correlation length is defined by the corresponding 
correlation function. For example, a particular important case of correlation functions 
is represented by reduced density matrices \cite{Coleman_29}. The latter can be treated 
as matrix elements of the reduced density operators $\hat{\rho}_n$, where $n = 1,2,\ldots$. 
The norm of a density operator, $||\hat{\rho}_n||$, is connected with its largest eigenvalue, 
which, in turn, is related to the effective correlation length. The trace of the density 
operator is a function of the number of particles $N$. For instance, $\rm{Tr} \rho_1 = N$. 
Then the exponent $\omega$, characterizing the relation between the norm and trace of a 
density operator, describes the dependence of the density operator spectrum on the number 
of particles, $||\hat{\rho}_1|| = N^{\omega}$. The parameter $\omega$ is an {\it order index} 
showing how quickly the order grows with the increase of the system size. When the order 
index is small, $\omega \ll 1$, this implies that the correlation length is much shorter
that the system effective length, hence there is no long-range order. But when the order 
index is close to one, then the correlation length becomes comparable to the system 
effective size, which means that long-range order is getting established in the system. 

It is the aim of the present paper to formulate these ideas, employing reduced density 
matrices for finite systems, in a rigorous quantitative way, to extend the consideration 
to general correlation operators, and to illustrate the approach by several concrete cases
of arising long-range order in the process of growing system size tending from finite $N$
to the thermodynamic limit. We consider the cases of Bose-Einstein condensation, 
superconducting transition, magnetic transition, and crystal-liquid transition, which, 
to make the ideas transparent, are treated in mean-field approximation.            
 
Throughout the paper, we use the system of units, where the Planck constant $\hbar$ and 
the Boltzmann constant $k_B$ are set to one.

\section{Operator order indices}

In the thermodynamic limit, the level of order of macroscopic systems can be characterized
by order indices of reduced density matrices \cite{Coleman_26,Coleman_27,Yukalov_28}. Here 
we extend the notion of order indices of reduced density matrices to finite statistical 
systems and generalize this notion to arbitrary correlation operators, showing that this 
notion provides a direct tool for measuring the level of order increasing together with 
the system size on the way to the thermodynamic limit.  
 
Order indices can be introduced for any trace-class operator, for which
$$
 0 < |{\rm Tr}\hat A| < \infty \; .
$$
In what follows, we shall deal with semi-positive operators. Let a trace-class semi-positive 
operator $\hat{A}$ be defined on a Hilbert space $\mathcal{H}$. The operator order index is
\be
\label{27}
 \om(\hat A) \; \equiv \; \frac{\log||\hat A||}{\log|{\rm Tr}\hat A| } \;  .
\ee 
Here the trace is taken over a Hilbert space $\mathcal{H}$ and the logarithm can be taken 
with respect to any base. Below, we prefer to use the natural logarithm. This index shows 
how the operator norm depends on the operator trace,
$$
|| \; \hat A \; || \; = \; |\; {\rm Tr}\; \hat A \; |^{\;\om(\hat A)} \;   .
$$ 
For semi-positive operators,
\be
\label{28}
|| \; \hat A \; || \; \leq \; {\rm Tr}\; \hat A \qquad ( \hat A \geq 0 ) \;  ,
\ee
then the order index is limited from above by one,
\be
\label{29}
\om(\hat A) \; \leq \; 1 \qquad (\hat A \geq 0 ) \;   .
\ee
From below, the index is not necessarily limited. There is order in the operator, when 
$\omega(\hat{A}) > 0$, while there is no order, if $\omega(\hat{A}) \leq 0$. Increasing
order index implies the growth of order. 

In general, the norm can be introduced in different ways. The operator norm
\be
\label{30}
|| \; \hat A \; || \; = \; \sup_\vp \; \frac{||\hat A\vp||}{||\vp||}
 \qquad ( ||\vp || \neq 0 ) \;  ,
\ee
where
$$
||\; \hat A \vp\; ||  \; = \; 
\sqrt{ \lgl \; \vp \; | \; \hat A^+  \hat A \; | \; \vp \; \rgl } \;  ,
$$
seems to be the most appropriate for the following applications. For Hermitian operators,
the operator norm reduces to the Hermitian norm
\be
\label{31}
||\; \hat A \; ||  \; = \; 
\sup_\vp \; \frac{\lgl \; \vp\; | \; \hat A\; | \; \vp\; \rgl}{||\vp||} 
\qquad ( \hat A^+ =  \hat A ) \;   .
\ee
This norm will be used throughout the paper, since we shall deal with Hermitian operators.  

Note that semi-positive trace-class operators are bounded, since  
\be
\label{B}
|| \; \hat A \; || \; \leq \; |{\rm Tr}  \; \hat A| < \infty \; .
\ee
Hence the order indices are well defined for this class of operators.

\section{Reduced density operators}

The notion of order indices, introduced above for arbitrary operators, can describe 
the ordering processes in physical systems when this notion is applied to the operators 
characterizing mutual correlations in these systems. The detailed information on the 
properties of statistical systems is contained in the reduced density matrices 
\cite{Coleman_29}. These matrices can be treated as matrix elements of reduced density 
operators.  

Thus, the first-order reduced density matrix, defined by the statistical average
\be
\label{32}
\rho(x,x') \; = \; {\rm Tr} \; \psi(x) \; \hat\rho \; \psi^\dgr(x') \; = \;
\lgl \; \psi^\dgr(x') \; \psi(x) \; \rgl \;   ,
\ee
where $x$ is a set of physical variables, like spatial coordinates, spin etc, can be treated 
as a matrix element of the first-order reduced density operator
\be
\label{33}
\hat\rho_1 \; = \; [\; \rho(x,x') \; ] \;   .
\ee
This operator is defined on the Hilbert space that is the closed linear envelope 
\be
\label{34}
\cH_1 \; = \; \overline\cL\{ \; | \; k \; \rgl \; \}
\ee
over the basis formed by natural orbitals 
\be
\label{35}
| \; k \; \rgl \; = \; | \; \vp_k \; \rgl \; = \; [\; \vp_k(x) \; ]
\ee
reflecting the system properties \cite{Coleman_29}, with $k$ being a set of quantum 
numbers labeling the orbitals.    

The norm of the first-order density operator is
\be
\label{36}
 ||\; \hat\rho_1\; || \; = \; \sup_k \; N_k \;  ,
\ee
where 
\be
\label{37}
N_k \; \equiv \; \lgl \; k \; | \; \hat\rho_1 \; | \; k \; \rgl \; = \;
\int \vp_k^*(x) \; \rho(x,x') \; \vp_k(x') \; dx dx'
\ee
plays the role of the occupation number associated with the state labeled by $k$. The 
trace of operator (\ref{33}) over the space (\ref{34}) is
\be
\label{38}
{\rm Tr}\; \hat\rho_1 \; = \; 
\sum_k  \lgl \; k \; | \; \hat\rho_1 \; | \; k \; \rgl \; = \; 
\int \rho(x,x) \; dx \; = \; N \; .
\ee
Then the order index for this operator, as defined in Ref. (\ref{27}), is 
\be
\label{39}
\omega(\hat\rho_1) \; = \; \frac{\log\; \sup_k N_k}{\log N} \; .
\ee
As is seen, the order index shows how quickly the largest occupation number increases
under the growing system size comprising $N$ particles,
$$
\sup_k N_k \; = \; N^{\om(\hat\rho_1)} \;   .
$$

Similarly, the second-order reduced density operator
\be
\label{40}
\hat\rho_2 \; = \; [\; \rho_2(x_1,x_2,x_1',x_2') \; ] 
\ee
is a matrix of the elements 
\be
\label{41}
\rho_2(x_1,x_2,x_1',x_2') \; = \; \lgl \; \psi^\dgr(x_2') \; \psi^\dgr(x_1') \;
\psi(x_1) \psi(x_2) \; \rgl  
\ee
enjoying the property
\be
\label{42}
 \rho_2(x_1,x_2,x_1',x_2') \; = \; \rho_2(x_2,x_1,x_2',x_1') \;  .
\ee
This operator acts on the Hilbert space
\be
\label{43}
 \cH^2 \; = \; \cH_1 \bigotimes \cH_2 \; = \; 
\overline\cL\{ \; | \; k p \; \rgl \; \}  
\ee
that is the closed linear envelope over the basis formed by the functions
\be
\label{44}
  | \; k p \; \rgl \; = \; | \; \vp_k \; \rgl \; \bigotimes \;
| \; \vp_p \; \rgl \;  = \; [\; \vp_k(x) \; ] \; \bigotimes \;
[\; \vp_p(x) \; ] \; .
\ee

The norm of the second-order operator (\ref{40}) reads as
\be
\label{45}
 || \; \hat\rho_2 \; || \; = \; \sup_{kp} \; N_{kp} \;  ,
\ee
where
$$
 N_{kp} \; = \; \lgl \; k p \; | \; \hat\rho_2 \; | \; kp \; \rgl \; = 
$$
\be
\label{46}
= \; 
\int \vp_k^*(x_1) \; \vp_p^*(x_2) \; \rho_2(x_1,x_2,x_1',x_2') \;
 \vp_k(x_1') \; \vp_p(x_2') \; dx_1 dx_2 dx_1' dx_2' \;  .
\ee
The trace over space (\ref{43}) is
\be
\label{47}
 {\rm Tr}\; \hat\rho_2 \; = \; 
\sum_{kp} \lgl \; kp \; | \; \hat\rho_2 \; | \; kp \; \rgl \; = \; 
\int \rho_2(x_1,x_2,x_1,x_2) \; dx_1 dx_2 \;  .
\ee
Then the order index of the second-order density operator is defined as
\be
\label{48}
 \om(\hat\rho_2) \; = \; \frac{\log\sup_{kp} N_{kp}}{\log{\rm Tr}\hat\rho_2} \;  .
\ee

In the same way, it is possible to define the $n$-th order reduced density operators
\be
\label{49}
 \hat\rho_n \; = \; [\; \rho_n(x_1,x_2,\ldots,x_n,x_1',x_2',\ldots,x_n') \; ] \;  ,
\ee
with the matrix elements
$$
\rho_n(x_1,x_2,\ldots,x_n,x_1',x_2',\ldots,x_n') \; = 
$$
\be
\label{50}
= \;
\lgl \; \psi^\dgr(x_n') \; \psi^\dgr(x_{n-1}') \ldots \psi^\dgr(x_1') \;
\psi(x_1) \; \psi(x_2) \ldots \psi(x_n) \; \rgl \;   ,
\ee
acting on the Hilbert space
\be
\label{51}
\cH^n \; = \; 
\cH_1 \bigotimes \cH_1  \bigotimes \cH_1 \bigotimes \ldots  \bigotimes \cH_1 \;  .
\ee
Then, it is straightforward to find the related order index $\omega(\hat{\rho}_n)$. 
However, the reduced density matrices of order higher than two are rarely considered,
since the main information is usually taken into account in the first-order and 
second-order reduced density matrices.

It is important to emphasize that the order indices for reduced density operators of any 
order are well defined for an arbitrary statistical system with a finite number of 
particles $N$. This follows from the known properties of the reduced density matrices 
\cite{Coleman_29}, for which the traces and norms of the related reduced density operators 
are finite for a finite number $N$. This also directly follows from the fact that the 
density operators are semi-positive trace-class operators for which the inequality (\ref{B})
is valid.

\section{Correlation operators}

Now we extend the notion of order indices to correlation operators that characterize a
kind of order in a system and which can be treated as matrix elements of operators. 
For example, let us have a two-point correlation function 
\be
\label{52}
C_1(x,x') \; = \; \lgl \; \hat A^+(x') \; \hat A(x) \; \rgl 
\ee
composed of some local observables $\hat{A}(x)$. Then it can be considered as a matrix 
element of the first-order correlation operator
\be
\label{53}
\hat C_1 \; = \; [\; C_1(x,x') \; ]   
\ee
acting on the Hilbert space $\mathcal{H}_1$ composed similarly to Ref. (\ref{34}). The 
first-order index for the correlation operator (\ref{53}) is given by 
\be
\label{54}
\om(\hat C_1) \; = \; \frac{\log||\hat C_1||}{\log|{\rm Tr}\;\hat C_1| } \; .
\ee

A second-order correlation operator   
\be
\label{55}
\hat C_2 \; = \; [\; C_2(x_1,x_2,x_1',x_2') \; ]  
\ee
can be defined as being composed of the matrix elements 
\be
\label{56}
C_2(x_1,x_2,x_1',x_2') \;  = \; 
\lgl \; \hat A^+(x_2') \; \hat A(x_1') \; \hat A(x_1) \; \hat A(x_2) \; \rgl \; .
\ee
The second-order operator acts on the Hilbert space $\mathcal{H}^2$ similar to space 
(\ref{43}). The order index for the correlation operator (\ref{55}) is
\be
\label{57}
\om(\hat C_2) \; = \; \frac{\log||\hat C_2||}{\log|{\rm Tr}\;\hat C_2| } \;   .
\ee

The introduced order indices are the functions of the system parameters and size, so 
that varying these parameters and the number of particles it is possible to study the 
level of ordering in the system.  

Here, for generality, we shall consider quantum systems. However, correlation functions
can be defined for quantum as well as for classical systems. As soon as a correlation
function $C_n(x_1,x_2,\ldots,x_n,x_1',x_2',\ldots,x_n')$ is given, independently 
of its origin, whether quantum or classical, the correlation operator $\hat C_n$ can be
defined as above.

\section{Bose-Einstein condensation}

Let us illustrate the behavior of the order indices by concrete examples. We start 
with Bose-Einstein condensation that in the recent years has been widely studied both 
theoretically and experimentally, as can be inferred from review articles and books
\cite{Courteille_31,Andersen_32,Yukalov_33,Bongs_34,Lieb_20,Yukalov_35,Posazhennikova_36,
Yukalov_21,Yukalov_19,Proukakis_37,Yurovsky_38,Yukalov_22,Yukalov_5,Yukalov_23,Yukalov_39,
Yukalov_2025,Castin_2025}. Here we study the dependence of order indices on the increasing 
system size. 

The Hamiltonian of a system of spinless particles is
\be
\label{58}
 \hat H \; = \; 
\int \psi^\dgr(\br) \; \left( - \; \frac{\nabla^2}{2m} \right) \; \psi(\br) \; d\br +  
\frac{1}{2} \int \psi^\dgr(\br) \; \psi^\dgr(\br') \; 
\Phi(\br-\br') \; \psi(\br') \; \psi(\br) \; d\br d\br' \; ,
\ee
where $\Phi$ is the interaction potential. This potential, for a dilute system, can be 
taken in the local form
\be
\label{59}
\Phi(\br) \; = \; \Phi_0 \; \dlt(\br) \; , \qquad 
\Phi_0 \; = \; 4\pi\; \frac{a_s}{m} \;   ,
\ee 
in which $a_s$ is scattering length, assumed to be positive, and $m$, particle mass.  

The possible appearance of Bose-Einstein condensate requires global gauge symmetry 
breaking, which is the necessary and sufficient condition for Bose condensation 
\cite{Lieb_20,Yukalov_21,Yukalov_22,Yukalov_2025}, and which can be realized by the 
Bogolubov shift 
$\psi(\br) \; = \; \eta(\br) + \psi_1(\br)$, where
\be
\label{60}
 \lgl \; \psi_1(\br) \; \rgl \; = \; 0 \; , \qquad
\int \eta^*(\br) \; \psi_1(\br) \; d\br \; = \; 0 \; .
\ee
The number of condensed and non-condensed particles, respectively, are 
\be
\label{61}
N_0 \; = \; \int |\; \eta(\br)\; |^2 d\br \; , \qquad
N_1 \; = \; \lgl \; \hat N_1 \; \rgl \;   ,
\ee
with the operator of non-condensed particles
\be
\label{62}
 \hat N_1 \; = \; 
\int  \psi_1^\dgr(\br) \; \psi_1(\br) \;d\br \;  .
\ee
The grand Hamiltonian, taking into account the normalizations (\ref{61}), is
\be
\label{63}
H   \; = \; \hat H - \mu_0 N_0 - \mu_1 \hat N_1 \;  .
\ee

Under the Bogolubov shift, there appear the expressions
\be
\label{64}
\rho_1(\br,\br') \; \equiv \; \lgl \; \psi_1^\dgr(\br') \; \psi_1(\br)  \; \rgl \; ,
\qquad
\sgm_1(\br,\br') \; \equiv \; \lgl \; \psi_1(\br') \; \psi_1(\br)  \; \rgl \; ,
\ee
called the normal and anomalous averages, respectively. The first-order reduced density 
matrix becomes
\be
\label{65}
 \rho(\br,\br') \; = \; \eta^*(\br') \; \eta(\br) + \rho_1(\br,\br') \;  .
\ee

For a uniform system, one has
\be
\label{66}
\eta \; = \; \sqrt{\rho_0} \; , \qquad
\rho_0 \; \equiv \; \frac{N_0}{V} \;   .
\ee
The operators of non-condensed particles can be expanded in plane waves,
\be
\label{67}
\psi_1(\br) \; = \; \sum_{k\neq 0} a_k \; \vp_k(\br) \; , \qquad
\vp_k(\br) \; = \; \frac{1}{\sqrt{V}} \; e^{i\bk\cdot\br} \;   .
\ee
The normal and anomalous averages take the form
\be
\label{68}
\rho_1(\br,\br') \; = \;  
\frac{1}{V} \sum_{k\neq 0} n_k \; e^{i\bk\cdot(\br-\br')} \; , 
\qquad 
\sgm_1(\br,\br') \; = \;  
\frac{1}{V} \sum_{k\neq 0} \sgm_k \; e^{i\bk\cdot(\br-\br')} \; ,
\ee
where
\be
\label{69}
n_k \; = \; \lgl \; a_k^+ \; a_k \; \rgl \; , \qquad
 \sgm_k \; = \; \lgl \; a_{-k} \; a_k \; \rgl \;  .
\ee
  
We employ the self-consistent mean-field approach that is gapless and conserving 
\cite{Yukalov_40,Yukalov_41,Yukalov_42,Yukalov_43}, and gives good quantitative 
agreement with Monte Carlo simulations for uniform Bose gas 
\cite{Giorgini_45,Pilati_46,Rossi_47} and with Monte Carlo results \cite{DuBois_48,DuBois_49}
for nonuniform trapped gas \cite{Yukalov_JPB}. In this approximation, we have 
\be
\label{70}
 n_k \; = \; 
\frac{\om_k}{2\ep_k} \; \coth\left( \frac{\ep_k}{2T}\right) - \; \frac{1}{2} \; ,
\qquad 
 \sgm_k \; = \; 
- \; \frac{mc^2}{2\ep_k} \; \coth\left( \frac{\ep_k}{2T}\right)  \; ,
\ee
where $T$ is temperature, the notation
\be
\label{71}
\om_k \; = \; m c^2 + \frac{k^2}{2m}
\ee
is used, and the spectrum of collective excitations is 
\be
\label{72}
\ep_k   \; = \; \sqrt{(ck)^2+\left( \frac{k^2}{2m}\right)^2} \;  .
\ee
The sound velocity $c$ is defined by the equation
\be
\label{73}
 mc^2  \; = \; \Phi_0 (\rho_0 + \sgm_1) \;  .
\ee
The density of non-condensed particles and the summary anomalous average are
\be
\label{74}
 \rho_1  \; = \; \frac{1}{V} \sum_{k\neq 0} n_k \; , 
\qquad
\sgm_1  \; = \; \frac{1}{V} \sum_{k\neq 0} \sgm_k \; ,
\ee
the total particle density being $\rho = \rho_0 + \rho_1$. 

Let us consider the case of zero temperature, $T=0$. Then
\be
\label{75}
 n_k \; = \; \frac{\om_k-\ep_k}{2\ep_k} \; , \qquad
\sgm_k \; = \; - \; \frac{mc^2}{2\ep_k} \qquad ( T = 0 ) \;  .
\ee
The density of non-condensed particles is
\be
\label{76}
\rho_1 \; = \; \frac{(mc)^3}{3\pi^2} \qquad (T = 0) \;   .
\ee
The anomalous average takes the form
\be
\label{77}
 \sgm_1 \; = \; - mc^2 \int \frac{1}{2\ep_k} \; \frac{d\bk}{(2\pi)^3} \qquad
(T = 0) \;  .
\ee

It is convenient to use the dimensionless quantities introducing the particle fractions
\be
\label{78}
n_0 \; \equiv \; \frac{\rho_0}{\rho} \; , \qquad
n_1 \; \equiv \; \frac{\rho_1}{\rho} \; , \qquad
\sgm \; \equiv \; \frac{\sgm_1}{\rho} \; , 
\ee
and the dimensionless sound velocities
\be
\label{79}
s \; \equiv \; \frac{mc}{\rho^{1/3}} \; , \qquad
s_B \; \equiv \; \frac{mc_B}{\rho^{1/3}}  \qquad 
\left( c_B \equiv \sqrt{\frac{\rho}{m}\; \Phi_0} \right) \;  .
\ee

The anomalous average (\ref{77}), because of the use of the contact potential, diverges 
and requires regularization. We employ the dimensional regularization 
\cite{Yukalov_23,Andersen_32} that is exact under asymptotically weak interactions, and 
can be analytically continued to finite interaction strength \cite{Yukalov_43}, which 
gives    
\be
\label{80}
\sgm \; = \; 
\frac{s_B^3}{\pi^2} \left( n_0 + \frac{s_B^3}{\pi^2} \; \sqrt{n_0} \right)^{1/2} \;   .
\ee
The interaction strength is characterized by the gas parameter
\be
\label{81}
\gm \; \equiv \; \rho^{1/3} \; a_s \;   ,
\ee
so that $s_B^2 = 4 \pi \gamma$. 

Finally, we obtain the closed set of equations for the fractions of condensed, $n_0$, 
and non-condensed, $n_1$, particles,
\be
\label{82}
 n_0 \; = \; 1 - \; \frac{s^3}{3\pi^2} \; , \qquad
n_1 \; = \; \frac{s^3}{3\pi^2} \;  ,
\ee
the anomalous average
\be
\label{83}
\sgm \; = \; \frac{8}{\sqrt{\pi} } \; \gm^{3/2} \left( n_0 +
 \frac{8}{\sqrt{\pi} } \; \gm^{3/2} \; \sqrt{n_0} \right)^{1/2} \;  ,
\ee  
and the sound velocity squared,
\be
\label{84}
 s^2 \; = \; 4\pi\gm ( n_0 + \sgm ) \;  .
\ee

These equations allow us to define the condensate fraction, the fraction of uncondensed
particles, the anomalous average, and the sound velocity as functions of the gas parameter.
Then the order indices, being the functions of these variables, can also be defined as 
functions of $\gamma$. In order to understand the influence of interactions on the order
in the system, we can find the explicit behavior of the solutions to the system of 
equations (\ref{82}), (\ref{83}), and (\ref{84}) for small gas parameters:
$$
n_0 \; \simeq \; 1 - \; 
\frac{8}{3\sqrt{\pi}} \; \gm^{3/2} - \; \frac{64}{3\pi}\; \gm^3 - \; 
\frac{256}{9\pi^{3/2}} \; \gm^{9/2} \; ,
$$
$$
\sgm \; \simeq \; 
\frac{8}{\sqrt{\pi}} \; \gm^{3/2} + \frac{64}{3\pi}\; \gm^3 - \; 
\frac{1408}{9\pi^{3/2}} \; \gm^{9/2} \; ,
$$
\be
\label{85}
s \; \simeq \; \sqrt{4\pi \gm} + \frac{16}{3} \; \gm^2 - \; 
\frac{64}{9\sqrt{\pi}} \; \gm^{7/2} \qquad (\gm \ra 0 ) \; .
\ee
For arbitrary gas parameters, the system of equations can be solved numerically, which is 
shown in Fig. 1.

\begin{figure}[ht]
\centerline{\includegraphics[width=10cm]{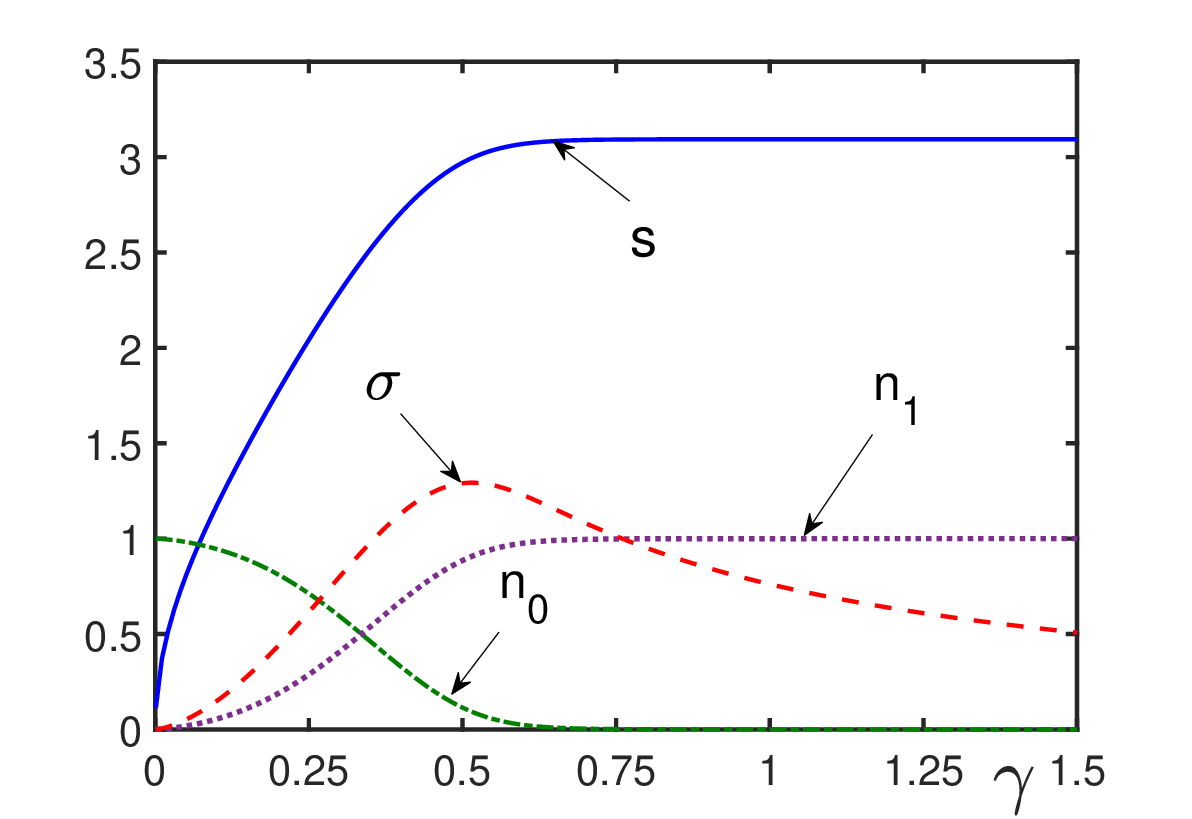}}
\caption{\small 
Condensate fraction $n_0$ (dashed-dotted line), fraction of uncondensed particles 
$n_1$ (dotted line), anomalous average $\sgm$ (dashed line), and dimensionless sound 
velocity $s$ (solid line) as functions of the gas parameter $\gm$, at zero temperature.
}
\label{fig:Fig.1}
\end{figure}

The order indices of reduced density operators are calculated as is described above. For 
the first-order density operator, the occupation numbers are
\be
\label{86}
N_k \; = \; N_0 \; \dlt_{k0} + n_k \;   .
\ee
For $n_k$ and $\sigma_k$ in the case of a finite system, there exists the minimal wave vector
\be
\label{87}
k_{min} \; = \; \frac{2\pi}{L} \; = \; \frac{2\pi}{V^{1/3}} \; = \; 
2\pi \; \left( \frac{\rho}{N}\right)^{1/3} \;   .
\ee
The norm of the first-order density operator reads as
\be
\label{88}
 ||\; \hat\rho_1\; || \; = \; 
\sup\left\{ \; n_0 N ; \; \frac{s}{4\pi}\;  N^{1/3} \; \right\} \;  ,
\ee
which results in the order index
\be
\label{89}
 \om(\hat\rho_1) \; = \; 
\frac{\ln \; \sup\left\{ \; n_0 N ; \; \frac{s}{4\pi}\;  N^{1/3} \; \right\} }{\ln N} \;  .
\ee

It is evident that for a macroscopic system in the thermodynamic limit, the whole system
is ordered,
\be
\label{90}
 \om(\hat\rho_1) \; \simeq \; 1 \qquad ( N \ra  \infty) \;  .
\ee

The order does not appear suddenly, but it grows with increasing system size. The order
index (\ref{89}) is calculated numerically using the solutions to the system of 
equations (\ref{82}), (\ref{83}), and (\ref{84}). The behavior of the order index for
different gas parameters is presented in Fig. 2 as a function of $\ln N$. The increase
of order depends on the value of the gas parameter. Thus for $\gamma = 0.1$ there is 
no order for small $N$, where the order index is negative, and the order quickly grows 
with $N$. For $\gamma = 0.5$, there is no order before $N = 9$, and the order index 
reaches $1/2$ close to $N = 77$. 

The interactions suppress the level of order diminishing the order index, which is clearly 
seen for small $\gamma$, where
\be
\label{91}
 \om(\hat\rho_1) \; \simeq \; 1 - \; \frac{8}{3\sqrt{\pi}\; \ln N} \; \gm^{3/2} -
\frac{224}{9\pi\; \ln N}\; \gm^3 \qquad ( \gm \ra 0 ) \;   .
\ee

\begin{figure}[ht]
\centerline{\includegraphics[width=10cm]{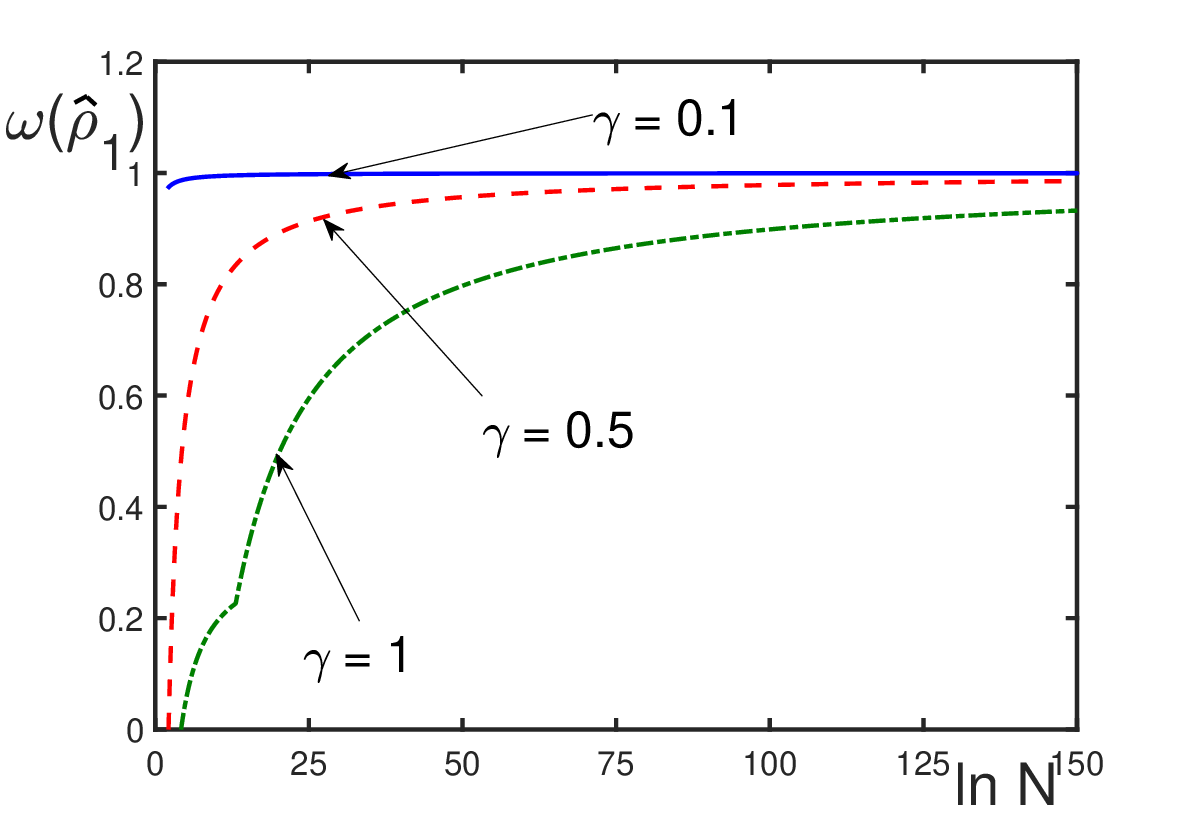}}
\caption{\small 
Appearance of order at Bose-Einstein condensation under increasing system size.
First order index $\om(\hat\rho_1)$ as a function of $\ln N$ for different gas 
parameters $\gm=0.1$ (solid line), $\gm=0.5$ (dashed line), and 
$\gm =1$ (dashed-dotted line).
}
\label{fig:Fig.2}
\end{figure}
 
The second-order reduced density matrix, taking account of the Bogolubov shift, and using 
the Hartree-Fock-Bogolubov (HFB) approximation, becomes
$$
\rho_2(\br_1,\br_2,\br_1',\br_2') \; = \; \rho_0^2 \; +
$$
$$
 + \;
\rho_0 \; [ \; \rho_1(\br_1,\br_1') +  \rho_1(\br_1,\br_2') +
\rho_1(\br_2,\br_1') +  \rho_1(\br_2,\br_2') + \sgm_1(\br_1,\br_2) + \sgm_1^*(\br_1',\br_2') \; ] +
$$
\be
\label{92}
+
\rho_1(\br_1,\br_1')\;  \rho_1(\br_2,\br_2') + \rho_1(\br_1,\br_2')\;  \rho_1(\br_2,\br_1')
+ \sgm_1^*(\br_1',\br_2')\; \sgm_1(\br_1,\br_2) \; .
\ee
The occupation numbers (\ref{46}) read as 
\be
\label{93}
N_{kp} \; = \; N_0^2 \; \dlt_{k0} \; \dlt_{p0} + N_0 \; ( n_k \; \dlt_{p0} +
n_p \; \dlt_{k0} ) + n_k \; n_p + 
\left( n_k^2 + |\; \sgm_k\; |^2 \right) \; \dlt_{kp} \;   .
\ee
Using for the terms $n_k$ and $|\sigma_k|$ expressions (\ref{75}), with the minimal 
wave vector (\ref{87}), we obtain the norm
\be
\label{94}
||\; \hat\rho_2 \; || \; = \; \sup \left\{ n_0^2 N^2 ; \; \frac{n_0 s}{4\pi}\; N^{4/3} ;
\; \frac{3 s^2}{16\pi^2}\; N^{2/3} \right\} \;   .
\ee
The second-order index reads as
\be
\label{95}
\om(\hat\rho_2) \; = \; 
\frac{\ln \; \sup \left\{ n_0^2 N^2 ; \; \frac{1}{4\pi}\; n_0 s N^{4/3} ;
\; \frac{3}{16\pi^2}\; s^2 N^{2/3} \right\} } {2\ln N} \;   .
\ee

In the thermodynamic limit, we have
\be
\label{96}
\om(\hat\rho_2) \; \simeq \; 1 \qquad ( N \ra \infty) \;   ,
\ee
while the behavior of the second-order index for finite values of $N$ is shown in 
Fig. 3. Again, we see that the growth of the order index depends on the interaction 
strength. Strong interactions make the development of order, under the increasing system 
size, more difficult. 

\begin{figure}[ht]
\centerline{\includegraphics[width=10cm]{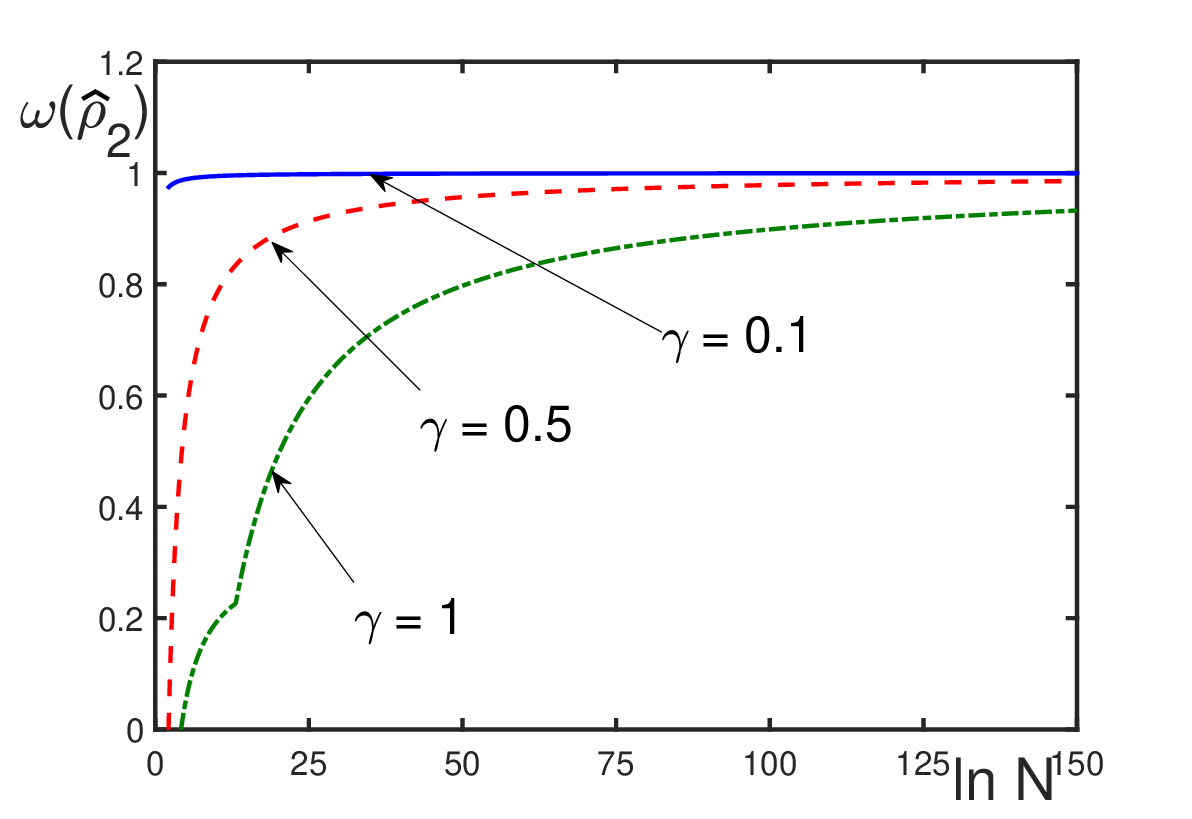}}
\caption{\small 
Appearance of order at Bose-Einstein condensation under increasing system size.
Second order index $\om(\hat\rho_2)$ as a function of $\ln N$ for different gas 
parameters $\gm=0.1$ (solid line), $\gm=0.5$ (dashed line), and 
$\gm=1$ (dashed-dotted line).
}
\label{fig:Fig.3}
\end{figure}

\section{Superconducting transition}

Starting from the general Hamiltonian with attractive interactions,
$$
H \; = \; \sum_s \int \psi_s^\dgr(\br) \; \left( - \; 
\frac{\nabla^2}{2m} - \mu \right) \; \psi_s(\br) \; d\br \; +
$$
\be
\label{97}
+ \;
\frac{1}{2} \sum_{s s'} \int  \psi_s^\dgr(\br) \;  \psi_{s'}^\dgr(\br') \;  
\Phi(\br-\br') \; \psi_{s'}(\br') \; \psi_s(\br) \; d\br d\br' \; ,
\ee
where the index $s$ labels spin variables, we pass to the momentum representation, 
assuming a uniform system, hence expanding the field operators over the plane waves,
\be
\label{98}
\psi_s(\br) \; = \; \frac{1}{\sqrt{V} } \sum_k c_{sk} \; e^{i\bk\cdot\br} \; .
\ee
Employing the Bardeen-Cooper-Schrieffer (BCS) restriction \cite{Bardeen_50}
\be
\label{99}
 c_{sk} \; c_{s' p} \; = \; 
\dlt_{-s s'} \; \dlt_{-k p} \; c_{sk} \; c_{-s,-k} \;  ,
\ee
one comes to the BCS Hamiltonian
\be
\label{100}
 H \; = \; 
\sum_s \; \sum_k \left( \frac{k^2}{2m} - \mu \right) \; c_{sk}^\dgr \; c_{sk} + 
\frac{1}{2V} \sum_s \; \sum_{kp} 
\Phi_{k-p} \; c_{sk}^\dgr \; c_{-s, -k}^\dgr \; c_{-s,-p} \; c_{sp} \; .
\ee
For the interaction term, the Hartree-Fock-Bogolubov approximation yields
$$
c_{sk}^\dgr \; c_{-s,-k}^\dgr \; c_{-s,-p} \; c_{sp} \; = \; \dlt_{kp}
\left( n_k \; c_{sk}^\dgr \; c_{sk} + n_k \; c_{-s,-k}^\dgr \; c_{-s,-k} -
n_k^2 \right) \; +
$$
\be
\label{101}
+ \;
\sgm_p \; c_{sk}^\dgr \; c_{-s,-k}^\dgr + 
\sgm_k \; c_{-s,-p}\; c_{sp} - \sgm_k^*\; \sgm_p \;  ,   
\ee
where the spin and momentum conservation in the binary statistical averages are taken 
into account, 
$$
\lgl \; c_{sk}^\dgr \; c_{s' p} \; \rgl \; = \; \dlt_{s s'} \; \dlt_{kp} \; n_k \; , 
\qquad
n_k \; \equiv \; \lgl \; c_{sk}^\dgr \; c_{s k} \; \rgl \; ,
$$
\be
\label{102}
\lgl \; c_{s'p} \; c_{s k} \; \rgl \; = \; \dlt_{-s s'} \; \dlt_{-kp} \; \sgm_k \; , 
\qquad
\sgm_k \; \equiv \; \lgl \; c_{-s,-k} \; c_{s k} \; \rgl \;  .
\ee
Thus we get the Hamiltonian
$$
H \; =\; \sum_s \; \sum_k \left( \frac{k^2}{2m} + \overline\Phi \; n_k - \mu \right) \;
c_{sk}^\dgr \; c_{sk} + 
$$
$$
+
\frac{1}{2V} \sum_s \; \sum_{kp} \Phi_{k-p} \left( \sgm_p\; c_{sk}^\dgr \; c_{-s,-k} +
\sgm_k^* \; c_{-s,-p} \; c_{sp} \right) -
$$
\be
\label{103}
- \; 
\frac{1}{2V} \sum_s \; \sum_{kp} \Phi_{k-p} \left( \sgm_k^* \; \sgm_p + \dlt_{kp} \; n_k^2 \right) \;  ,
\ee
where
\be
\label{104}
\overline\Phi \; \equiv \;  \frac{1}{V} \int \Phi(\br) \; d\br \;  .
\ee

To diagonalize the Hamiltonian, one resorts to the Bogolubov canonical transformation
\be
\label{105}
c_{sk} \; = \; u_k \; b_{sk} + v_k \; b_{-s,-k}^\dgr \;   ,
\ee
in which
$$
|\; u_k\; |^2 + |\; v_k\; |^2 \; = \; 1 \; , \qquad 
u_k \; v_{-k} + u_{-k} \; v_k \; = \; 0 \; ,
$$
$$
|\; u_k\; |^2 \; = \; \frac{1}{2} \; \left( 1 + \frac{\om_k}{\ep_k} \right) \; ,
\qquad
|\; v_k\; |^2 \; = \; \frac{1}{2} \; \left( 1 - \; \frac{\om_k}{\ep_k} \right) \; .
$$
Here the notation
\be
\label{106}
\om_k \; \equiv \; \frac{k^2}{2m} + \overline\Phi \; n_k - \mu 
\ee
is used. The spectrum of collective excitations has the form
\be
\label{107}
\ep_k \; = \; \sqrt{\Dlt_k^2 + \om_k^2} \;   ,
\ee
where the gap is
\be
\label{108}
\Dlt_k \; = \; -\; \frac{1}{2V} \sum_p \Phi_{k-p} \; \sgm_p \;  .
\ee
Then we get the diagonal Hamiltonian
\be
\label{109}
H \; = \; \sum_s \; \sum_k \ep_k \; b_k^\dgr \; b_k +
\sum_k ( \om_k - \ep_k + \Dlt_k \; \sgm_k) \; ,
\ee
for which 
\be
\label{110}
 \lgl \; b_{sk}^\dgr \; b_{sk} \; \rgl \; = \; \frac{1}{\exp(\ep_k/T)-1} \;  .
\ee

The averages (\ref{102}) become
\be
\label{111}
n_k \; = \; \frac{1}{2} - \; 
\frac{\om_k}{2\ep_k} \; \tanh\left( \frac{\ep_k}{2T}\right) \; ,
\qquad
\sgm_k \; = \;  \frac{\Dlt_k}{2\ep_k} \; \tanh\left( \frac{\ep_k}{2T}\right) \;    .
\ee
The chemical potential is defined by the equation
\be
\label{112}
\frac{2}{V} \sum_k n_k \; = \; 2 \int n_k \; \frac{d\bk}{(2\pi)^3} \; = \; \rho \; .
\ee

To find the first-order index, we need to consider the first-order reduced density 
operator
\be
\label{113}
 \hat\rho_1 \; = \; [ \; \rho(\br,\br') \; ] \;  ,
\ee
in which the density matrix is
\be
\label{114}
\rho(\br,\br') \; = \; \lgl \; \psi_s^\dgr(\br') \; \psi_s(\br) \; \rgl \; = \;
\frac{1}{V} \sum_k n_k \; e^{i\bk\cdot(\br-\br')} \;  .
\ee
Then we have
\be
\label{115}
  ||\; \hat\rho_1 \; || \; = \; \sup_k n_k \; , \qquad
{\rm Tr} \; \hat\rho_1 \; = \; N \; .
\ee
Hence the first-order index is
\be
\label{116}
\om(\hat\rho_1) \;  = \; \frac{\ln \; \sup_k n_k}{\ln N} \;   .
\ee
Since $\sup n_k \leq 1/2$, we see that the first order index
\be
\label{117}
\om(\hat\rho_1) \;  \simeq \; 0 \qquad ( N \gg 1)
\ee
does not show order.

The second-order reduced density operator
\be
\label{118}
 \hat\rho_2 \; = \; [\; \rho(\br_1,\br_2,\br_1',\br_2') \; ]
\ee
is expressed through the matrix elements
\be
\label{119}
\rho(\br_1,\br_2,\br_1',\br_2') \; = \; \lgl \; \psi_{-s}^\dgr(\br_2') \; \psi_s^\dgr(\br_1') \;
\psi_s(\br_1) \; \psi_{-s}(\br_2) \; \rgl \;   ,
\ee
which in the HFB approximation read as
\be
\label{120}
\rho(\br_1,\br_2,\br_1',\br_2') \; = \; \rho(\br_1,\br_1') \; \rho(\br_2,\br_2') +
\sgm^*(\br_2',\br_1') \; \sgm(\br_2,\br_1) \;  ,
\ee
where the anomalous average is
\be
\label{121}
\sgm(\br,\br') \; = \; \lgl \; \psi_{-s}(\br') \; \psi_s(\br) \; \rgl \; = \;
\frac{1}{V} \sum_k \sgm_k \; e^{i\bk\cdot(\br-\br')} \;  .
\ee

The anomalous average defines the number of pairs of correlated particles by the 
expression
\be
\label{122}
 N_\pi \; = \; \int |\; \sgm(\br_2,\br_1) \; |^2 d\br_1 d\br_2 \; = \; 
\sum_k |\; \sgm_k\; |^2 \;  .
\ee
Hence the fraction of pair-correlated particles is
\be
\label{123}
  n_{cor} \; = \; \frac{2 N_\pi}{N} \; .
\ee
The trace of the density operator (\ref{118}) takes the form
\be
\label{124}
 {\rm Tr} \; \hat\rho_2 \; = \; 
\int \rho(\br_1,\br_2,\br_1,\br_2) \; d\br_1 d\br_2 \; = \; 
N^2 + \frac{1}{2} \; n_{cor} \; N \;  .
\ee

In the presence of the pair-correlated particles, the eigenvalue problem
\be
\label{125}
\int \rho(\br_1,\br_2,\br_1',\br_2') \;
\vp(\br_1',\br_2') \; d\br_1 d\br_2 \; = \; 
\lbd \; \vp(\br_1,\br_2)
\ee
gives the normalized eigenfunction
\be
\label{126}
\vp(\br_1,\br_2) \; = \; \frac{\sgm(\br_2,\br_1)}{\sqrt{N_\pi}}
\ee
and the eigenvalue
\be
\label{127}
 \lbd \; = \; \frac{1}{N_\pi} \sum_k n_k^2 \; |\; \sgm_k \; |^2 + N_\pi \;  .
\ee

As is seen from expressions (\ref{111}), both $n_k$ and $|\sigma_k|$ quickly diminish
with increasing $k$, so that 
$$
 \sum_k n_k^2 \; |\; \sgm_k \; |^2 \; \cong \; 
\sup_k n_k^2 \sum_k |\; \sgm_k\; |^2 \;  .
$$
Then Eq. (\ref{127}) can be written as 
\be
\label{128}
 \lbd \; = \; \sup_k n_k^2 + N_\pi \;  .
\ee
Taking into account that $\sup_k n_k = 1/2$, we have 
\be
\label{129}
||\; \hat\rho_2 \; || \; = \;
 \frac{1}{4} + \frac{1}{2} \; n_{cor} \; N \;   .
\ee

Thus we come to the second order index
\be
\label{130}
 \om(\hat\rho_2) \; = \; \frac{\ln\left( \frac{1}{4} + \frac{1}{2} \; n_{cor} N\right)}
{\ln\left( N^2 + \frac{1}{2} \; n_{cor} N\right)} \; .
\ee
This shows that in thermodynamic limit
\be
\label{131}
 \om(\hat\rho_2) \; \simeq \; \frac{1}{2} \qquad ( N \ra \infty) \;   ,
\ee
provided $n_{cor} > 0$. The behavior of the order index (\ref{130}) as a function of 
arbitrary $\ln N$ is presented in Fig. 4. The order grows fast with the increase of the 
system size. Even for a small fraction of correlated particles $n_{cor} = 0.1$, the order 
index becomes nonzero for $N > 15$ and reaches $1/4$ for $N = 404$. 

\begin{figure}[ht]
\centerline{\includegraphics[width=10cm]{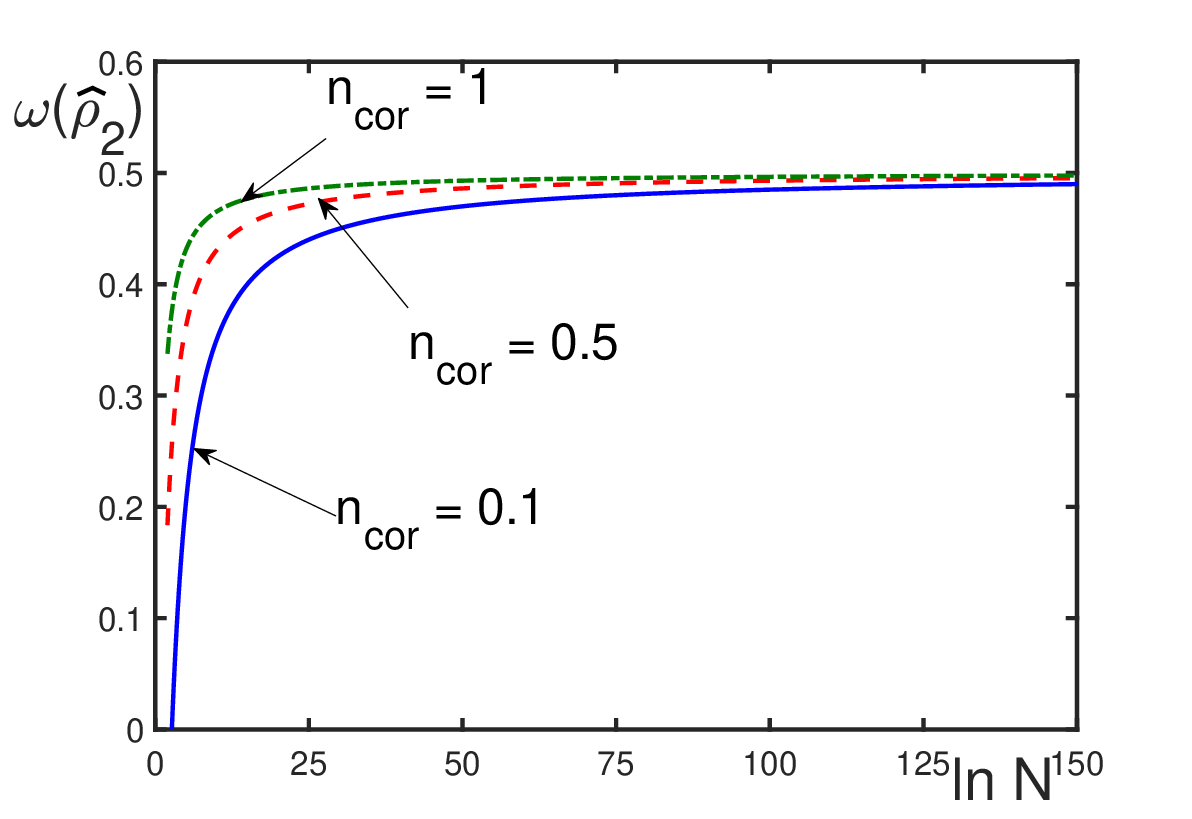}}
\caption{\small 
Appearance of order at superconduction transition under increasing system size.
Second order index $\om(\hat\rho_2)$ as a function of $\ln N$ for different 
fractions of pair-correlated particles $n_{cor}=0.1$ (solid line), $n_{cor}= 0.5$ 
(dashed line), and $n_{cor}=1$ (dashed-dotted line).
}
\label{fig:Fig.4}
\end{figure}

\section{Magnetic transition}

To describe how magnetic order grows with the increasing system size, it is necessary 
to consider the correlation operators composed of spin operators. Let us consider spin 
operators $S_i^z$ corresponding to spin one-half ($S = 1/2$). The matrix elements of the 
first-order correlation operator 
\be
\label{132}
\hat C_1 \; = \; [\; C_{ij} \; ]
\ee
are given by the correlation functions     
\be
\label{133}
C_{ij} \; = \; \lgl \; S_i^z \; S_j^z \; \rgl \;   .
\ee
The operator (\ref{132}) acts on the Hilbert space
\be
\label{134}
\cH_1 \; = \; \overline\cL \left\{ \vp_k(\ba_j) = 
\frac{1}{\sqrt{N}} \; e^{i\bk\cdot\ba_j} \right\}
\ee
that is a closed linear envelope over a complete orthonormal basis, such that
$$
\sum_{i=1}^N \vp_k^*(\ba_i) \; \vp_p(\ba_i) \; = \; \dlt_{kp} \; ,
\qquad
\sum_k\vp_k^*(\ba_i) \; \vp_k(\ba_j) \; = \; \dlt_{ij} \;   .
$$
For concreteness, we consider spin $S = 1/2$. Keeping in mind long-range interactions
between the spins, we use the mean-field approximation for different lattice sites,
\be
\label{135}
 \lgl \; S_i^z \; S_j^z \; \rgl \; = \; 
\lgl \; S_i^z \; \rgl \lgl \; S_j^z \; \rgl 
\qquad ( i \neq j) \; .
\ee
In that way, the correlation function is
\be
\label{136}
 \lgl \; S_i^z \; S_j^z \; \rgl \; = \; 
S^2 \; [ \; \dlt_{ij} + ( 1 - \dlt_{ij}) \; M^2 \; ] \;  ,
\ee
where the notation
\be
\label{137}
M \; \equiv \; \frac{1}{S} \; \lgl \; S_i^z \; \rgl
\ee
is used.

Employing the matrix element
\be
\label{138}
\lgl \; k \; | \; \hat C_1 \; | \; p \; \rgl \; = \;
\sum_{ij} \vp_k^*(\ba_i) \; C_{ij} \; \vp_p(\ba_j) \; = \;
S^2 \; [ \; 1 + \dlt_{k0} \; \dlt_{p0} \; M^2 ( N - 1) \; ] \;   ,
\ee  
we find the norm
\be
\label{139}
 ||\; \hat C_1 \; || \; = \; \sup_k \;
\lgl \; k \; | \; \hat C_1 \; | \; k \; \rgl \; = \;  S^2 \; [ \; 1+ M^2 (N - 1) \; ]
\ee
and the trace 
\be
\label{140}
{\rm Tr} \; \hat C_1 \; = \; \sum_{i=1}^N C_{ii} \; = \; S^2 N \;   .
\ee

Thus the first-order index is
\be
\label{141}
 \om(\hat C_1) \; = \; \frac{\ln\{ S^2[\;1+M^2(N-1)\;]\}}{\ln (S^2 N)} \;  .
\ee
In the thermodynamic limit, 
\be
\label{142}
 \om(\hat C_1) \; \simeq \; 1 \qquad ( N \ra \infty) \; ,   
\ee
as soon as $M > 0$. The behavior of the order index (\ref{141}) for arbitrary $N$
is shown in Fig. 5.

\begin{figure}[ht]
\centerline{\includegraphics[width=10cm]{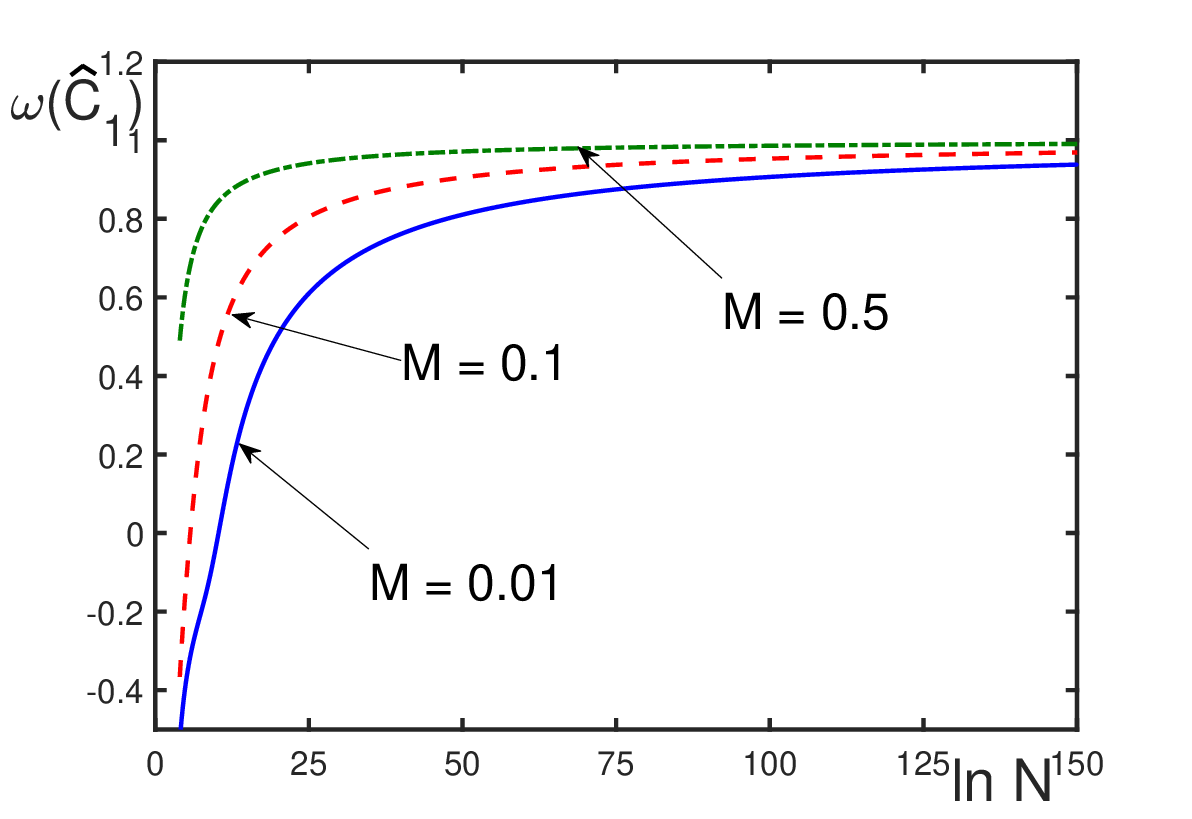}}
\caption{\small 
Appearance of order at magnetic transition under increasing system size.
First order index $\om(\hat C_1)$ as a function of $\ln N$ for different 
magnetizations $M \equiv (1/S) \lgl S_i^z \rgl$, with $M=0.01$ (solid line), 
$M= 0.1$ (dashed line), and $M=0.5$ (dashed-dotted line).
}
\label{fig:Fig.5}
\end{figure}

The second-order correlation operator
\be
\label{143}
\hat C_2 \; = \; [ \; C_{ijmn} \; ]
\ee
is composed of the correlation functions
\be
\label{144}
 C_{ijmn} \; \equiv \; \lgl \; S_i^z \; S_j^z \; S_m^z \; S_n^z \; \rgl \; .
\ee
Invoking the same procedure as above, with the use of the binary Ter Haar 
decoupling, we have
$$
  C_{ijmn} \; = \; S^4 \; \left\{ \dlt_{ij} \; \dlt_{mn} +
\dlt_{ij} ( 1 - \dlt_{mn})\;  M^2 +
 ( 1 - \dlt_{ij})\;  \dlt_{mn} \; M^2 + \right.
$$
\be
\label{145}
\left. +
 ( 1 - \dlt_{ij})\; ( 1 - \dlt_{mn}) \; M^4 \right\} \;  .
\ee
With the matrix element 
$$
\lgl \;  kp \; | \; \hat C_2 \; | \; kp \; \rgl \; = \; S^4 \left\{
\dlt_{-kp} + 2 (\dlt_{k0}\; \dlt_{p0} \; N - \dlt_{-kp}) \; M^2 \right. +
$$
\be
\label{146}
+ \left.
( \dlt_{k0}\; \dlt_{p0} \; N^2 - 2 \dlt_{k0}\; \dlt_{p0} \; N +  \dlt_{-kp}) \; M^4
\right\} \;  ,
\ee
we obtain the norm
\be
\label{147}
 ||\; \hat C_2 \; || \; = \; \sup_{kp} 
\lgl \; kp \; | \; \hat C_2 \; | \; kp \; \rgl \; = \;
S^4 \; [ \; 1 + ( N - 1) M^2 \; ]^2 \;  .
\ee
The trace of the correlation operator (\ref{143}) is
\be
\label{148}
 {\rm Tr} \; \hat C_2 \; = \; \sum_{ij} C_{ijij} \; = \; S^4 N^2 \;  .
\ee
Therefore the second-order index coincides with the first,
\be
\label{149}
 \om(\hat C_2) \; = \; \om(\hat C_1) \;  .
\ee

\section{Crystal-liquid transition}

The notion of order indices can be used for describing different types of arising order,
whether off-diagonal or diagonal. As an illustration, let us show how the crystal order,
arising under crystallization process can be treated. The starting point is the definition 
of appropriate correlation operators. 

Let us consider the density operator
\be
\label{150}
 \hat\rho(\br) \; \equiv \; \psi^\dgr(\br) \; \psi(\br) \;  ,
\ee
whose statistical average gives the local particle density in the real space,
\be
\label{151}
\rho(\br) \; = \; \lgl \; \hat\rho(\br) \; \rgl \;   .
\ee
The average density over the system is
\be
\label{152}
  \rho \; = \; \frac{1}{V} \int \rho(\br) \; d\br \; = \; \frac{N}{V} \; .
\ee
Let us label the locations of the maximal density by the vectors ${\bf a}_j$, with 
$j=1,2,\ldots,N$, so that
\be
\label{153}
 \max_{\br} \rho(\br) \; = \; \rho(\ba_j) \;  .
\ee
The fraction operator, associated with the points of the maxima, is given by the expression
\be
\label{154}
 \hat n_j \; \equiv \; \frac{\hat\rho(\ba_j)}{\rho} \;  .
\ee
  
And let us define the deviation operator 
\be
\label{155}  
\hat D_j \; \equiv \; \hat n_j -1
\ee
characterizing the deviations of the local maximal density from the average density, 
so that
\be
\label{156}
 D \; \equiv \; \lgl \; \hat D_j \; \rgl \; = \; 
\lgl \; \hat n_j \; \rgl - 1 \; = \; 
\lgl \; \frac{\hat\rho(\ba_j)}{\rho} \; \rgl - 1 \; .
\ee

Assuming long-range interactions allowing for the mean-field approximation
\be
\label{157}
\lgl \; \hat n_i \; \hat n_j \; \rgl \; = \;
\lgl \; \hat n_i  \; \rgl  \lgl \; \hat n_j \; \rgl \qquad ( i \neq j ) \; ,
\ee
we have
\be
\label{158}
\lgl \; \hat D_i \; \hat D_j \; \rgl \; = \;
\lgl \; \hat D_i  \; \rgl  \lgl \; \hat D_j \; \rgl \; \equiv \; D^2 
\qquad ( i \neq j ) \;    .
\ee
For a liquid, with a uniform density, $D = 0$, while for a crystal, whose density 
is nonuniform, $D > 0$. 

We define the correlation operator
\be
\label{159}
 \hat C_1 \; = \; [ \; C_{ij} \; ] \; , \qquad
C_{ij} \; \equiv \;  \lgl \; \hat D_i \; \hat D_j \; \rgl \; ,
\ee
whose matrix elements are the correlation functions corresponding to the deviation 
operators. This operator acts on the Hilbert space (\ref{134}).

With the matrix elements 
\be
\label{160}
\lgl \; k \; | \; \hat C_1 \; | \;  k \; \rgl \; = \; 
\lgl \; \hat D_j^2 \; \rgl + \dlt_{k0} \; D^2 ( N - 1) \;  ,
\ee
we find the norm
\be
\label{161}
 || \; \hat C_1 \; || \; = \;  \lgl \; \hat D_j^2 \; \rgl + D^2 ( N - 1)  
\ee
and the trace
\be
\label{162}
 {\rm Tr} \; \hat C_1 \; = \; \lgl \; \hat D_j^2 \; \rgl \; N \;  .
\ee

Then we obtain the first-order index
\be
\label{163}
 \om(\hat C_1 ) \; = \; \frac{\ln[\; \lgl \; \hat D_j^2 \; \rgl +D^2(N-1)\;]}
{\ln(  \lgl \; \hat D_j^2 \; \rgl N) }
\ee
that is similar to the index (\ref{141}) characterizing the arising magnetic order.
In the thermodynamic limit,
\be
\label{164}
\om(\hat C_1 ) \; \simeq \; 1 \qquad ( N \ra \infty) \;   ,
\ee
provided that $D > 0$. The overall behavior of the order index (\ref{163}) with respect 
to the varying particle number $N$ is analogous to that of the magnetic order index 
(\ref{141}).

\section{Conclusion}

The process describing how the order grows with the increasing system size on the way to
the thermodynamic limit is analyzed. For this purpose, a sequence of systems, with fixed 
parameters, but increasing system size, such that the number of particles $N$ and volume 
$V$ increase, while the density $\rho = N/V$ kept fixed, is considered. Each system in the 
sequence is assumed to be an equilibrium statistical system, which requires that the number 
of its constituents be sufficiently large, $N \gg 1$, and the system lifetime be sufficiently 
long, longer than the observation time. In this setup, no nonequilibrium effects need to be 
considered.        

The arising and growing order, increasing together with the system size, can be quantified
by the order indices that can be introduced for arbitrary trace-class operators, such as
the reduced density operators, or other correlation operators. Of course, for quite small 
systems there could exist finite-size corrections. However, in small systems, where the 
order has not yet been developed, the order index is also quite small. Noticeable order 
emerges in sufficiently large systems with the number of particles $N \gg 1$, when boundary 
corrections do not play important role, and practically disappear for very large $N$, 
completely vanishing in the thermodynamic limit when $N \ra \infty$.    
 
With the increasing system size, hence the increasing number of particles $N$, the order 
index grows from zero or even negative, showing the absence of order, to finite positive 
values, reaching, in the thermodynamic limit, $N \ra \infty$ the maximal possible value. 
Thus, for the cases where the order is due to some arrangement of individual variables, 
the order index reaches the value one, while in the case, where the order is caused by 
pairing correlations, the single-particle order index shows no order and the second-order 
index can reach the value $1/2$. The order indices, increasing with the growing system 
size, are considered for the processes of Bose-Einstein condensation, superconductivity, 
magnetization, and crystallization.    

In conclusion, it is useful to emphasize once more why the introduced notion of order 
indices can describe the emergence and growth of order under the increasing system size
and to remind the qualitative physical picture behind the formulas.   

In a microscopic system of just a few particles there can be no order. When the system 
increases reaching mesoscopic size, where the number of particles $N$ is sufficiently 
large, there can appear a metastable ordered phase having a finite lifetime. More detailed
discussion of characteristic times is postponed to the Appendix. Here we concentrate on
spatial correlations that are in the origin of arising order.   

Correlation length $l_{cor}$ is connected with the reduced density matrix 
$\rho({\bf r},{\bf r}')$ and can be defined by the relation
\be
\label{165}
 \rho l_{cor}^3 \; = \; 
\sup_k \int \vp_k^*(\br) \; \rho(\br,\br') \; \vp_k(\br') \; d\br d\br' \;  ,
\ee
thus specifying the volume where the density matrix is essentially nonzero. By using the 
notation for the average distance between particles, $a$, given by the identity
\be
\label{166}
 \rho a^3  \; = \; 1 \; ,
\ee 
and noticing that the right-hand side of (\ref{165}) is the norm of $\hat{\rho}_1$, we have
\be
\label{167}
 \left( \frac{l_{cor}}{a} \right)^3 \; = \; ||\; \hat\rho_1 \; ||\;  .
\ee

In order to show that these equations really define correlation length, let us consider the
limiting cases. When the density matrix describes the ultimate short-range correlations,
so that $\rho({\bf r},{\bf r}') \sim \delta({\bf r}-{\bf r}')$, then $l_{cor} \sim a$.
In the opposite case, when there exist long-range correlations, such that 
$\rho({\bf r},{\bf r}') \sim \rho$ throughout the whole system of length $L \sim V^{1/3}$,
then (\ref{165}) gives $l_{cor} \sim L$. This behavior perfectly corresponds to physical 
intuition that the system is completely ordered when the correlation length is comparable
to the system length. In this way, the increase of the density operator norm leads to the
increase of the correlation length and to the rise of the order index. 

The correlation length $l_{cor}$ characterizes correlations between single particles. It may 
happen that single particles do not exhibit long-range correlations, but pairs of particles
do experience long-range correlations, as it happens for superconductors. This implies
that there exist, actually, different correlation lengths, with respect to single particles
and pairs of particles. The correlation length for pairs of particles is defined by the
equation
\be
\label{168}
 \left( \frac{l_{cor}^{(2)}}{a} \right)^6 \; = \; ||\; \hat\rho_2 \; ||\;  .
\ee
This correlation length can vary between the limiting cases of the average particle distance 
$a$ and the system length $L$.     

In the same way, it is straightforward to introduce the correlation length $l_{cor}^{(n)}$ 
characterizing correlations between the groups of $n$ particles,
\be
\label{169}
\left( \frac{l_{cor}^{(n)}}{a} \right)^{3n} \; = \; ||\; \hat\rho_n \; ||\;   .
\ee
Thence, the growth of a correlation length implies the increase of the operator norm and 
of the corresponding order index. In a similar way, correlation lengths can be defined for 
correlation operators of different nature, e.g., based on correlation functions for spin 
operators, 
\be
\label{170}
\left( \frac{l_{cor}^{(n)}}{a} \right)^{3n} \; = \; ||\; \hat C_n \; ||\;   .
\ee
 
As has been noticed, correlations and, respectively, the ordering can be absent for single 
particles but present for particle pairs, as in superconductors. General relations between
the existence of ordering in correlation operators depend on particular systems and concrete 
types of studied ordering. Below, we summarize our knowledge on the existence of ordering
and the behavior of order indices for correlation operators. For concreteness, we can keep
in mind the reduced density operators $\hat{\rho}_n$ that are a particular kind of correlation 
operators.

\vskip 2mm

{\it Conjecture 1}. If some kind of ordering happens in a system, there should exist a 
related correlation operator and a corresponding order index depending on the system size. 

\vskip 2mm 

{\it Conjecture 2}. If there is a long-range order in $\hat{\rho}_1$, it also exists in 
all higher orders of $\hat{\rho}_n$, with admissible $n=2,3,\ldots$. 

\vskip 2mm

{\it Conjecture 3}. If there is no order in $\hat{\rho}_1$, this does not mean its absence
in some higher orders of $\hat{\rho}_n$.

\vskip 2mm

{\it Conjecture 4}. If there is long-range order in $\hat{\rho}_2$, it exists in all 
even-orders of $\hat{\rho}_{2n}$, with admissible $n=2,3,\ldots$.   

\vskip 2mm
  
Effects of asymptotic symmetry breaking can be observed in different finite systems, such as 
trapped atoms, quantum dots, metallic grains, and atomic nuclei \cite{Birman_51}. The notion 
of order indices can be useful for the description of and manipulation with these finite 
quantum systems.

\vskip 3mm

{\bf Appendix}. {\it Characteristic times}

\vskip 2mm

To better understand the physics of processes occurring in finite statistical systems,
and the meaning of equilibrium in such systems, it is worth recalling several important 
facts.   

The notion of thermodynamic equilibrium presupposes that the system can be observed for 
a sufficiently long time, longer than the time of local fluctuations, so that what is 
observed and measured in an equilibrium system are its average properties  
\cite{Bogolubov_1,Bogolubov_2,Kardar_52,Bogolubov_3,Yukalov_4}. In an infinite 
system, an equilibrium thermodynamics phase, under appropriate conditions, can live 
infinite time. In a finite system, there are restrictions on the lifetime of a metastable 
state. 

First of all, there is a restriction from below, requiring that the lifetime be much longer 
than the local equilibration time \cite{Bogolubov_3, Yukalov_9,Yukalov_10}, otherwise no 
ordered phase can arise. This time can be estimated as $t_{loc} \sim \lambda/ v$, where 
$\lambda \sim 1/ (\rho r_{int}^2) \sim a^3/r_{int}$ is the mean free path, 
$v\sim \hbar /(m r_{int})$, characteristic particle velocity, $a$, mean interparticle 
distance, $r_{int}$, interaction radius, and $m$, particle mass. Thus the local equilibration 
time is $t_{loc} \sim m a^3/{\hbar r}_{int}$. For the parameters typical of condensed matter, 
the local equilibration time is of the order $10^{-13}$ s. 

In some cases, there can appear mesoscopic fluctuations involving several particles and living
longer than the local equilibration time. The size of such fluctuations is much larger than 
the mean interparticle distance $a$ that in condensed matter is usually close to the 
interaction radius. The mesoscopic fluctuations representing one phase inside another are 
called heterophase fluctuations \cite{Yukalov_9,Frenkel_7,Frenkel_8}. They can occur when their 
existence diminishes the system free energy. The typical lifetime of such fluctuations in 
condensed matter is $t_{mes} \sim 10^{-12}$ s. In the interval between $t_{loc}$ and $t_{mes}$,
the system has yet to be treated as nonequilibrium, becoming equilibrium for the times much 
longer than $t_{mes}$. 

Thus, in the situation, where there are no strongly nonequilibrium initial conditions, the 
system can be considered as equilibrium when the observation time is much longer then the 
local equilibration time $t_{loc}$ and the mesoscopic fluctuation time $t_{mes}$, while from 
above, the observation time is limited by the lifetime $t_{met}$ of the considered metastable 
state. In this way, a system can be handled as equilibrium if the observation time $t_{obs}$ 
is in the interval of times
$$
t_{loc} \ll t_{mes} \ll t_{obs} \ll t_{met}  \; .
$$

The lifetime of a metastable state can be estimated in the following way. Let the space 
of microscopic states of the considered system be a Hilbert space $\mathcal{H}$.
Different phases are characterized by different subspaces of the total system space. 
Say the subspace of typical disordered states is $\mathcal{H}_0$ and the subspace of 
typical ordered states is $\mathcal{H}_1$. The separation of the ordered and disordered 
states can be done by means of the method of restricted trace 
\cite{Yukalov_9,Yukalov_10,Brout_11,Rudnick_12,Palmer_13}. The subspaces $\mathcal{H}_0$ 
and $\mathcal{H}_1$ can be represented as weighted Hilbert spaces 
\cite{Kantorovich_14,Kolmogorov_15}, with the weights selecting the required typical 
states \cite{Yukalov_9,Yukalov_10}. 

Suppose the space $\mathcal{H}_0$ possesses a vacuum state $|0 \rangle$, while the space
$\mathcal{H}_1$ possesses a vacuum state $|0 \rangle_1$. All other states with the 
corresponding typical properties of the studied phases are formed by generating them from 
the corresponding vacuum states. The lifetime of a metastable phase can be estimated as
the escape time required for the transition from one phase to another \cite{Gardiner_16}.
The escape rate (or transition rate) reads as $\gamma_{esc} = p_{esc}/t_0$, where $p_{esc}$ 
is the escape probability $p_{esc} = |\langle 0|0\rangle_1|^2$ and $t_0$ is a 
characteristic time of the order of the local equilibration time. The rates are inversely 
proportional to the related times, so that $\gamma_{esc} \equiv 1/t_{esc}$. Thus the 
phase lifetime is given by the expression
\be
\label{3}
 t_{esc} \; = \; \frac{t_0}{|\; \lgl\; 0 \; | \; 0 \; \rgl_1\; |^2} \; .
\ee
The transition between different phases, characterized by different Hilbert spaces, is 
somewhat similar, although mathematics is of course rather different, to the transition 
between different parts of the same Hilbert space defining the localization time 
\cite{Cohen_17}. The other analogy is the transition (escape) from one potential well 
into another well in real space \cite{Gardiner_16}. 

To be more precise and to better understand how the expression for the escape time can 
be derived, let us illustrate this by a concrete example of a transition from the normal,
uncondensed, phase to the Bose-condensed phase. Mathematical details describing systems
with Bose-Einstein condensate can be found in Refs. \cite{Yukalov_19,Yukalov_18}.  

Let us consider a system of Bose particles characterized by the field operators 
$\psi({\bf r})$ satisfying Bose commutation relations. Strictly speaking, the field 
operator depends as well on time which we do not write explicitly for the sake of 
simplicity of notation. For the same reason, for a while we do not mention internal 
degrees of freedom, if any, which can be included by treating the field operators as 
columns with respect to these variables. We do not want to complicate the consideration, 
when it is not important, as far as our principal aim is to study the growth of order 
under the increase of the system size in real space. When necessary, we can include 
internal degrees of freedom, e.g. spin.  

Defining the vacuum state, such that
\be
\label{4}
 \psi(\br) \; | \; 0 \; \rgl \; = \; 0 \;  ,
\ee
it is possible to generate the states
\be 
\label{5}
|\; \vp\; \rgl \; = \; \sum_{n=0}^\infty \frac{1}{\sqrt{n!}}
\int\vp(\br_1,\br_2,\ldots,\br_n) \; 
\prod_{i=1}^n \psi^\dgr(\br_i) \; d\br_i \; | \; 0 \; \rgl \; ,
\ee
where $\varphi({\bf r}_1, {\bf r}_2, \ldots, {\bf r}_n)$ is a symmetric function of its 
arguments. The closed linear envelope of all states $|\varphi\rangle$ generated from the 
vacuum state $|0\rangle$ by the field operators $\psi$, forms the Fock space 
\be
\label{6}
 \cF(\psi) \; = \; \overline\cL \{ \; |\; \vp\; \rgl \; \} \;  .   
\ee

The standard system Hamiltonian, with pair interactions, is invariant with respect to 
the global gauge symmetry, when the operators $\psi({\bf r})$ are replaced by 
$\psi({\bf r}) e^{i \alpha}$, with a real phase $\alpha$.

If in the system there starts appearing Bose-Einstein condensate, its correct description
requires the occurrence of gauge symmetry breaking, since this breaking is the necessary 
and sufficient condition for Bose-Einstein condensation 
\cite{Lieb_20,Yukalov_21,Yukalov_22,Yukalov_5,Yukalov_23}. The most convenient way of gauge 
symmetry breaking is the use of the Bogolubov shift \cite{Bogolubov_1,Bogolubov_2} that is 
an exact canonical transformation 
\be
\label{7}
\psi(\br) \; = \; \eta(\br) + \psi_1(\br) \;   .
\ee
Here the field operator $\psi_1$, satisfying the Bose commutation relations, is the field 
operator of uncondensed particles and $\eta$ is the condensate wave function normalized 
to the number of condensed particles
\be
\label{8}
N_0 \; = \; \int |\; \eta(\br) \; |^2 d\br \;   .
\ee
  
As is easy to notice, the state $|0\rangle$, which is a vacuum with respect to $\psi$, 
is not a vacuum with respect to $\psi_1$, since
\be
\label{9}
\psi_1(\br) \; | \; 0 \; \rgl \; = \; - \eta(\br) \; | \; 0 \; \rgl 
\ee
is not zero. Hence, with respect to $\psi_1$, there exists another vacuum defined by
the condition
\be
\label{10}
 \psi_1(\br) \; | \; 0 \; \rgl_1 \; = \; 0 \; .
\ee
From this vacuum, it is straightforward to generate the functions
\be
\label{11}
|\; \vp_1 \; \rgl \; = \; 
\sum_{n=0}^\infty \frac{1}{\sqrt{n!} } \int \vp_1(\br_1,\br_2,\ldots,\br_n) \; 
\prod_{n=1}^n \psi_1^\dgr(\br) \; d\br_i \; | \; 0 \; \rgl_1
\ee
forming the Fock space as their closed linear envelope 
\be
\label{12}
\cF(\psi_1) \; = \; \overline\cL\{ \; | \; \vp_1\; \rgl \; \} \; .
\ee
        
Canonical transformations can be realized by transformation operators \cite{Beresin_24}. 
For example, the Bogolubov shift (\ref{7}) can be represented as due to the action of
the Bogolubov operator
\be
\label{13}
\hat B \; = \; \exp(\hat D) \; , \qquad 
\hat B^{-1} \; = \; \exp(-\hat D) \;   ,
\ee
in which 
\be
\label{14}
 \hat D \; = \; 
\int [\; \eta^*(\br) \; \psi(\br) - \eta(\br) \; \psi^\dgr(\br) \;] \; d\br \; .    
\ee
The Bogolubov shift (\ref{7}) is equivalent to the transformation
\be
\label{15}
\psi(\br) \; = \; \hat B \; \psi_1(\br) \; \hat B^{-1}  \; ,
\qquad 
\psi_1(\br) \; = \; \hat B^{-1} \; \psi(\br) \; \hat B  \; .
\ee
Respectively, the vacuum states are related by the transformations 
\be
\label{16}
|\; 0 \; \rgl \; = \; \hat B \; |\; 0 \; \rgl_1 \; , \qquad
 |\; 0 \; \rgl_1 \; = \; \hat B^{-1} |\; 0 \; \rgl \;   .
\ee

For two operators $\hat{A}_1$ and $\hat{A}_2$, whose commutator $[\hat{A}_1,\hat{A}_2]$
commutes with both $\hat{A}_1$ and $\hat{A}_2$, the Baker-Hausdorff formula is valid,
$$
 \exp(\hat A_1 + \hat A_2) \; = \;
e^{\hat A_1} \; e^{\hat A_2} \; 
\exp\left( - \; \frac{1}{2} \; [\; \hat A_1 , \; \hat A_2\; ] \right) \;  .
$$ 
Using this, we have
\be
\label{17}
\hat B^{-1} \; = \; \exp\left\{ \int \eta(\br) \; \psi^\dgr(\br) \; d\br \right\} \;
\exp\left\{ - \int \eta^*(\br) \; \psi(\br) \; d\br \right\}  \;
\exp\left\{ - \; \frac{1}{2} \int |\; \eta(\br) \; |^2 \; d\br \right\} \; .
\ee
Acting by the operator (\ref{17}) on the vacuum state $|0\rangle$, we find
\be
\label{18}       
\hat B^{-1} |\; 0 \; \rgl \; = \; 
\exp\left\{ - \; \frac{1}{2} \int |\; \eta(\br) \; |^2 \; d\br \right\} \;
\exp\left\{ \int \eta(\br) \; \psi^\dgr(\br) \; d\br \right\} \; | \; 0 \; \rgl \; .
\ee

Here it is useful to remember that a coherent state that is the solution of the equation 
\be
\label{19}
 \psi(\br) \; | \; \eta \; \rgl \; = \; \eta(\br) \; | \; \eta \; \rgl \; ,
\ee 
where $\eta({\bf r})$ is a coherent field \cite{Yukalov_5}, has the form
\be
\label{20}
 |\; \eta \; \rgl \; = \;  
\exp\left\{ - \; \frac{1}{2} \int |\; \eta(\br) \; |^2 \; d\br \right\} \;
\exp\left\{ \int \eta(\br) \; \psi^\dgr(\br) \; d\br \right\} \; | \; 0 \; \rgl \;   
\ee
coinciding with the right-hand side of Eq. (\ref{18}). Hence the vacuum $|0\rangle_1$ 
is a normalized coherent state,
\be
\label{21}
|\; 0 \; \rgl_1 \; = \; \hat B^{-1} \; |\; 0 \; \rgl \; = \;
|\; \eta \; \rgl \; .
\ee
This is in agreement with the equation
\be
\label{22}
\psi(\br) \; |\; 0 \; \rgl_1 \; = \; \eta(\br) \; |\; 0 \; \rgl_1
\ee
following from the form (\ref{7}) of the Bogolubov shift.    

Taking into account normalization (\ref{8}), we obtain
\be
\label{23}
\lgl \; 0 \; | \; 0 \; \rgl_1 \; = \; \exp\left( - \frac{1}{2} \; N_0 \right) \;   .
\ee
Then the transition probability is
\be
\label{24}
|\; \lgl \; 0 \; | \; 0 \; \rgl_1 \; |^2 \; = \; \exp( - N_0 ) \;   .
\ee
Therefore the escape time (\ref{3}) reads as
\be
\label{25}
 t_{esc} \; = \; t_0 \; e^{N_0} \;  .
\ee

Formula (\ref{25}) shows that if the system is small and the ordered phase is also small, 
such that $N_0$ is not much larger than one, it is not stable and, occasionally appearing, 
disappears very quickly. But in a larger system that can be sufficiently large, it becomes 
more stable and can live longer. Thus, with the increasing system size, the possibility 
of a larger ordered phase increases, which implies the growing order. In the thermodynamic 
limit, when $N_0 \propto N \ra \infty$, an absolutely equilibrium ordered phase can become
absolutely stable.      

It is useful to mention that, in general, the phase factor of a coherent state is not 
defined, so that there can appear the condensate with the coherent field $\eta e^{i \alpha}$ 
that is normalized to the same number $N_0$ as in the integral (\ref{8}), and with an 
arbitrary phase $\alpha$. The phase appears randomly. But after appearing, it is preserved 
as far as the transition to another condensate with a different phase $\alpha \neq 2 \pi n$, 
with $n = 0,1,2,\ldots$, is suppressed, since the transition rate between different 
condensates
$$
|\; \lgl \; \eta \; | \; \eta \; e^{i\al} \; \rgl \; |^2 \; = \;
\left| \; \exp\left\{ - N_0 \left( 1 - e^{i\al} \right) \right\} \; \right|^2 \; = \;
\exp\left\{ - 4N_0 \; \sin^2\left( \frac{\al}{2}\right) \right\}
$$
is close to zero for large $N_0$ and $\alpha \neq 2 \pi n$. Therefore the first randomly 
appearing condensate phase prevails and, without the loss of generality, it can be set 
to zero.  
     
The picture, delineated above for the appearing Bose condensate under the increase of 
the system size, remains similar for the arising and growing order of any kind. In a 
microscopic system, there is no yet order. In a mesoscopic system there arises the 
embryo of the ordered phase, which lives finite time $t_{esc}$. Then, under the appropriate 
conditions, with the growth of the system, the size of the ordered phase grows and is 
getting more and more stable. The escape time increases by the law
\be
\label{26}
 t_{esc} \; = \; t_0 \; e^{c N_0} \;  ,
\ee
where $t_0$ is of the order of the local equilibration time, which is the minimal time 
for establishing local equilibrium required for the ordered phase appearance, $c$ is a 
constant of order one, and $N_0$ is the number of particles in the ordered phase. 
Expression (\ref{26}), to some extent, reminds us the form of the Arrhenius law 
\cite{Laidler_25}, however the meaning is rather different. The Arrhenius law defines
the typical time (or rate) for the transition between locations in the real space, while 
here $t_{esc}$ defines the transition between Fock spaces. 

The typical lifetime $t_{esc}$, is estimated by the formula (\ref{26}), where $t_0$ is 
of the order of local equilibration time of $10^{-13}$ s and $c$ is a constant of order 
one, characterizing the system. As is seen, for the number of particles in the ordered 
phase $N_0$ of order one, the lifetime does not differ much from the local equilibration 
time, while for $N_0$ of order $10$, the lifetime can be of order $10^{-9}$ s. For the 
number of particles in the system of order $50$, the lifetime of the ordered phase 
reaches the order of $16$ years. With increasing $N$, the lifetime quickly grows to huge 
values. This explains why phase transitions and ordered phases that, strictly speaking, 
in finite systems are metastable, do exist and can be observed without any problem.

\vskip 3mm

{\bf Declarations}

\vskip 2mm

{\bf Author contributions}: All authors contributed to the study conception and design. 
Material preparation, data collection and analysis were performed by V.I. Yukalov and 
E.P. Yukalova. The first draft of the manuscript was written by V.I. Yukalov and all 
authors commented on previous versions of the manuscript. All authors read and approved 
the final manuscript.

\vskip 2mm

{\bf Funding}: No funding was received to assist with the preparation of this manuscript.

\vskip 2mm

{\bf Financial interests}: The authors have no relevant financial or non-financial 
interests to disclose. 

\vskip 2mm

{\bf Competing interests}: The authors have no competing interests to declare that are 
relevant to the content of this article.

\end{document}